\documentclass[3p,times,procedia]{elsarticle}
\usepackage{nupha_ecrc}

\volume{00}

\firstpage{1}

\journalname{Nuclear Physics A}

\runauth{}

\jid{nupha}

\jnltitlelogo{Nuclear Physics A}

\usepackage{amssymb}
\usepackage{amsfonts}
\usepackage{amssymb}
\usepackage[figuresright]{rotating}
\usepackage{xspace}
\usepackage{calrsfs}

\newcommand{\sqrts}{\sqrt{s}}
\newcommand{\pT}{\rm p_{T}}
\newcommand{\sqrtsnn}{\sqrt{s_{_{\mbox{\rm \tiny{NN}}}}}}

\newcommand{\Lint}{\mathcal{L}_{\mbox{\rm \tiny{int}}}}

\newcommand{\jpsi}    {\rm J/\psi}
\newcommand{\gaga}    {\gamma\,\gamma}
\newcommand{\qqbar}    {q\bar{q}}
\newcommand{\ccbar}    {c\bar{c}}
\newcommand{\bbbar}    {b\bar{b}}
\newcommand{\ttbar}    {t\bar{t}}

\newcommand{\epem}{\rm e^{+}\,e^{-}}

\newcommand\cO{{\cal O}}

\newcommand*{\ie}{i.e.\@\xspace}
\newcommand*{\cm}{c.m.\@\xspace}

\begin{document}

\begin{frontmatter}



\dochead{XXVIth International Conference on Ultrarelativistic Nucleus-Nucleus Collisions\\ (Quark Matter 2017)}

\title{Physics with ions at the Future Circular Collider}


\author{David~d'Enterria$^\dagger$ for the FCC-ions study group (L.~Apolinario, N.~Armesto, A.~Dainese, 
J.~Jowett, J.P.~Lansberg, S.~Masciocchi, G.~Milhano, 
C.~Roland, C.A.~Salgado, M.~Schaumann, M.~van Leeuwen, U.A.~Wiedemann)}

\address{$^\dagger$CERN, EP Department, 1211 Geneva, Switzerland}

\begin{abstract}
The unique physics opportunities accessible with nuclear collisions at the CERN Future Circular Collider (FCC) 
are summarized. Lead-lead (PbPb) and proton-lead (pPb) collisions at $\sqrtsnn$~=~39 and 63~TeV respectively
with $\Lint$~=~33~nb$^{-1}$ and 8~pb$^{-1}$ monthly integrated luminosities, will provide unprecedented 
experimental conditions to study quark-gluon matter at temperatures $\cO$(1~GeV). The following topics 
are succinctly discussed: 
(i) charm-quark densities thrice larger than at the LHC, leading to direct heavy-quark impact in the bulk QGP properties, 
(ii) quarkonia, including $\Upsilon(1S)$, melting at temperatures up to five times above the QCD critical temperature, 
(iii) access to initial-state nuclear parton distributions (nPDF) at fractional momenta as low as $x\approx 10^{-7}$, 
(iv) availability of about $5\cdot 10^5$ top-quark pairs per run to study the high-$x$ gluon nPDF and the 
energy loss properties of boosted colour-antennas, 
(v) study of possible Higgs boson suppression in the QGP, and
(vi) high-luminosity $\gaga$ (ultraperipheral) collisions at c.m. energies up to 1~TeV.
\end{abstract}


\end{frontmatter}



\section{Introduction}

The FCC is a future collider project under study at CERN aiming at proton-proton (pp) collisions at a centre-of-mass (\cm) 
energy $\sqrts$~=~100~TeV in a new 80--100 km tunnel with 16--20~T dipoles~\cite{Benedikt:2015poa}. 
The FCC operation with nuclear beams yields nucleon-nucleon \cm\ energies 
$\rm \sqrtsnn=\!100~TeV\cdot\sqrt{Z_1Z_2/(A_1A_2)}\!$~=~39~TeV (63~TeV) for PbPb (pPb), \ie\ $\times$7 
larger than those accessible at the LHC. The expected (monthly) integrated luminosities 
are also $\times$10 above the LHC values: $\Lint$~=~33~nb$^{-1}$ and 8~pb$^{-1}$ for PbPb and 
pPb~\cite{Schaumann:2015fsa,Dainese:2016gch}. 
This writeup succinctly summarizes the novel physics opportunities at reach with nuclear collisions at the 
FCC, covering the detailed studies presented in~\cite{Dainese:2016gch} plus some more recent developments.

\section{Global medium properties, and small-$x$ QCD}

\begin{figure*}[hbpt!]
\centering
\includegraphics[width=0.37\textwidth,height=6.5cm]{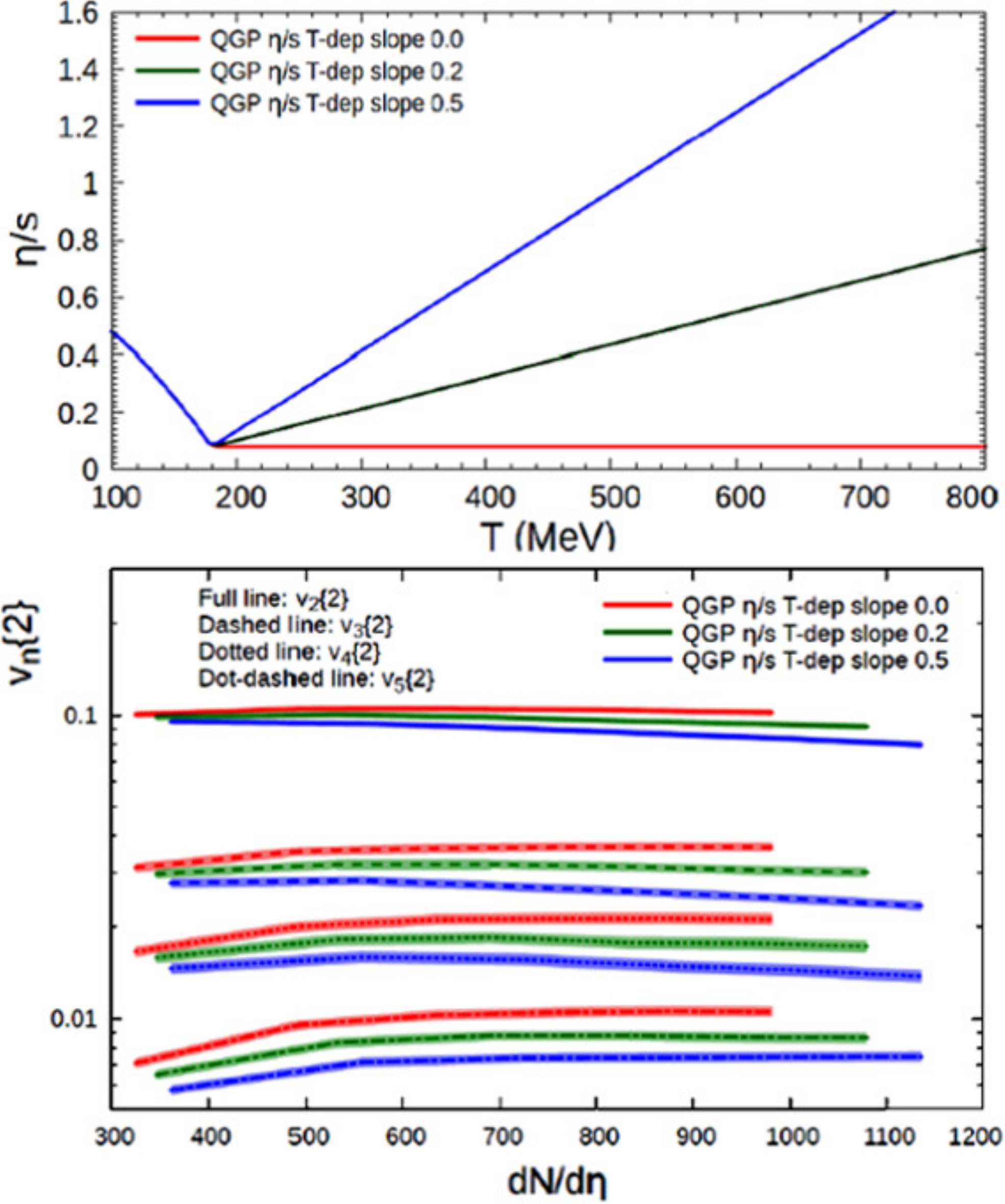}\hspace{0.5cm}
\includegraphics[width=0.47\textwidth,height=6.5cm]{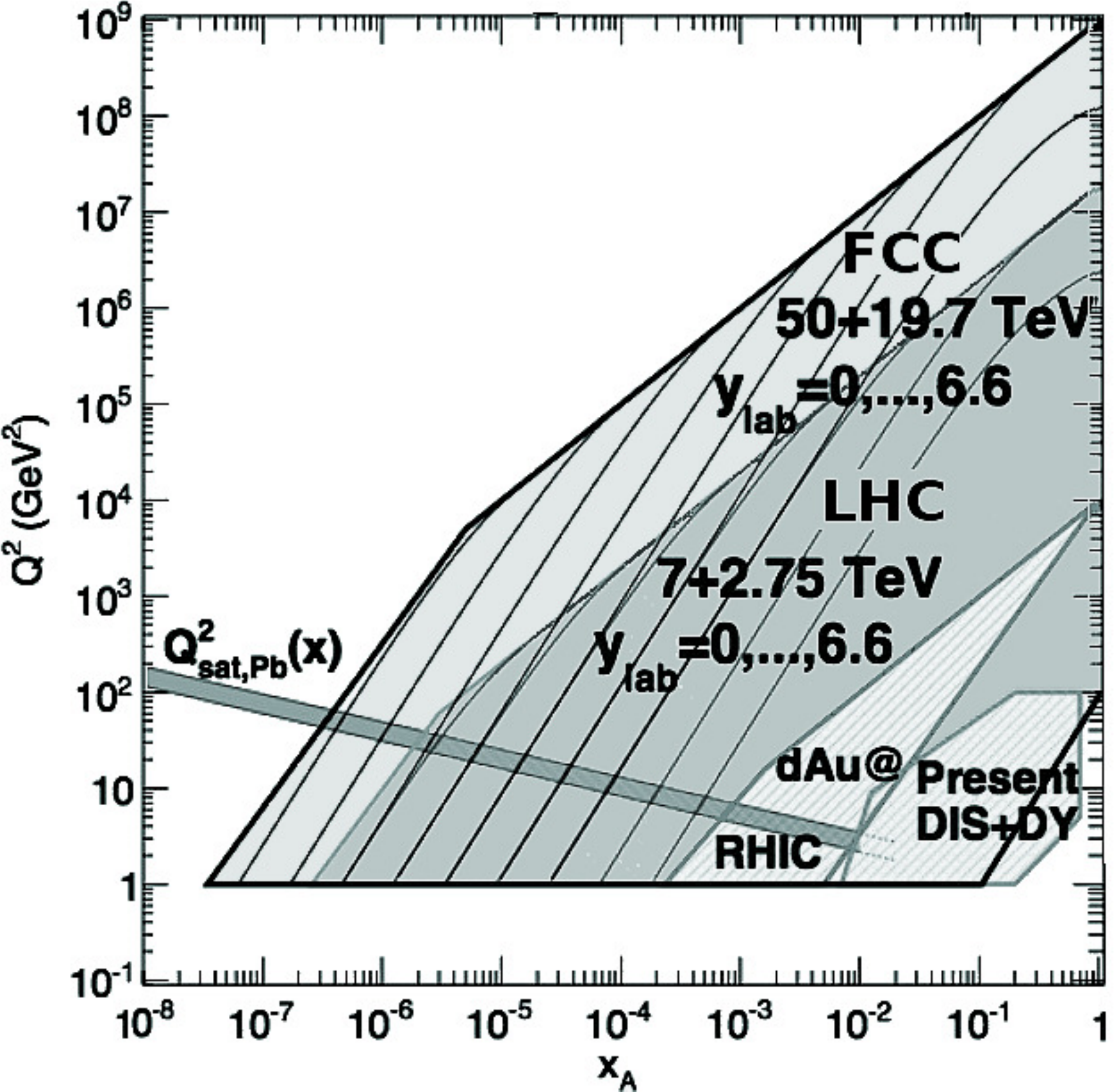}
\caption{Left: Different dependencies of the $\eta/s$ ratio versus temperature (top) and associated
viscous hydrodynamics predictions of the $v_n$ azimuthal coefficients as a function of particle density (bottom).
Right: Regions of the $x$--$Q^2$ plane covered in pPb at FCC and LHC
(for varying rapidities $y_{\rm lab}$~=~0--6.6) and in nuclear DIS+DY,
compared to the expected saturation scale $Q^2_{sat}(x)$ curve.}
\label{fig:1}
\end{figure*}

At 39~TeV, central PbPb collisions will produce $\rm dN_{ch}/d\eta|_{\eta=0}\approx$~3\,600 charged 
particles per unit rapidity at midrapidity, according to an extrapolation of the current data based on a
$\propto(\!\sqrtsnn)^{0.3}$  dependence. For a participant volume of V~=~11\,000~fm$^{3}$, the formation of 
a quark-gluon plasma (QGP) with 35--40~GeV/fm$^{3}$ energy density is expected at $\tau_0$~=~1~fm/c~\cite{Dainese:2016gch}.
At earlier times, the medium temperatures will reach T~$\approx$~1~GeV. 
The twofold larger multiplicities and hotter medium open up the possibility to carry out
event-by-event measurements of high-harmonics $v_n$ of the azimuthal distribution of particles 
with respect to the reaction plane over a large density range. Such Fourier coefficients 
are sensitive to different dependencies of the QGP shear-viscosity over entropy-density 
($\eta/s$) ratio as a function of temperature (Fig.~\ref{fig:1}, left).

The large FCC \cm\ energies and correspondingly broad rapidity ranges for particle 
production result in a very wide kinematical reach in terms of probed fractional 
momenta $x$ in the nuclear parton distribution functions (nPDF).
Semihard ($Q\approx$~5~GeV) forward ($|y|>3$) particles will probe nPDF down to 
$x \approx Q\,e^{-y}/\sqrtsnn\approx$~10$^{-7}$, while cross sections of the 
``standard candles'' W and Z bosons will be sensitive to $x\approx$~10$^{-4}$ at midrapidity (Fig.~\ref{fig:1}, right).
Since currently there are no direct data constraints on nPDF below $x\approx 5\cdot10^{-3}$~\cite{Eskola:2009uj},
the FCC data provides a huge lever-arm to study the small-$x$ parton structure and evolution.
Back-to-back jet/hadron correlations at forward $y$ in pPb~\cite{Rezaeian:2012ye}, and
exclusive $\jpsi$ production in ultraperipheral (photon-induced) PbPb collisions~\cite{Armesto:2017hrs}
will be processes particularly sensitive to the saturation regime (below the $Q_{sat}$ curve in Fig.~\ref{fig:1}, right), 
where standard DGLAP collinear factorization is expected to break down~\cite{Gelis:2010nm}.

\section{Open charm and bottom, quarkonia, and top quarks}

PbPb collisions at 39~TeV will lead to a very abundant production of charm quarks from both initial 
gluon-gluon interactions (the $\ccbar$ cross sections will even exceed $\rm \sigma_{inel}(pPb,PbPb)$, 
indicating a very large probability for multiple charm production per event~\cite{dEnterria:2016yhy})
plus secondary $gg\to\ccbar$ processes in the thermalizing plasma~\cite{Zhou:2016wbo,Liu:2016zle} which 
increase the final yield by 50--100\%. Such large charm densities will have two main
consequences. First, since finite-T QCD calculations~\cite{Laine:2006cp} indicate that charm quarks start 
contributing as thermal degrees-of-freedom in the QGP equation of state for T$\approx$~350~MeV,
they will modify the bulk thermodynamic properties of the medium compared to lower-energy collisions, 
as well as actively participate in the collectively flowing medium. Secondly, thermal 
$\ccbar$ quarks can recombine into $\jpsi$ bound states with larger probability than at the LHC~\cite{Andronic:2011yq}, 
leading to a change in the $\jpsi$ production pattern from the suppression 
observed so far to an enhanced production 
at low $\pT$ compared to pp collisions at the same energy~\cite{Liu:2016zle}.
In the bottom-quark sector, the initial QGP temperatures $\cO$(1~GeV) can lead to direct melting of the $\Upsilon$(1S)
bound-state, expected by lattice-QCD calculations to occur at T~=~(4--5)T$_c$~\cite{Aarts:2014cda}, although the large 
density of $\bbbar$ pairs may partially compensate such a suppression through recombination 
in the plasma~\cite{Andronic:2011yq}.

\begin{figure*}[hbpt!]
\centering
\includegraphics[width=0.50\textwidth]{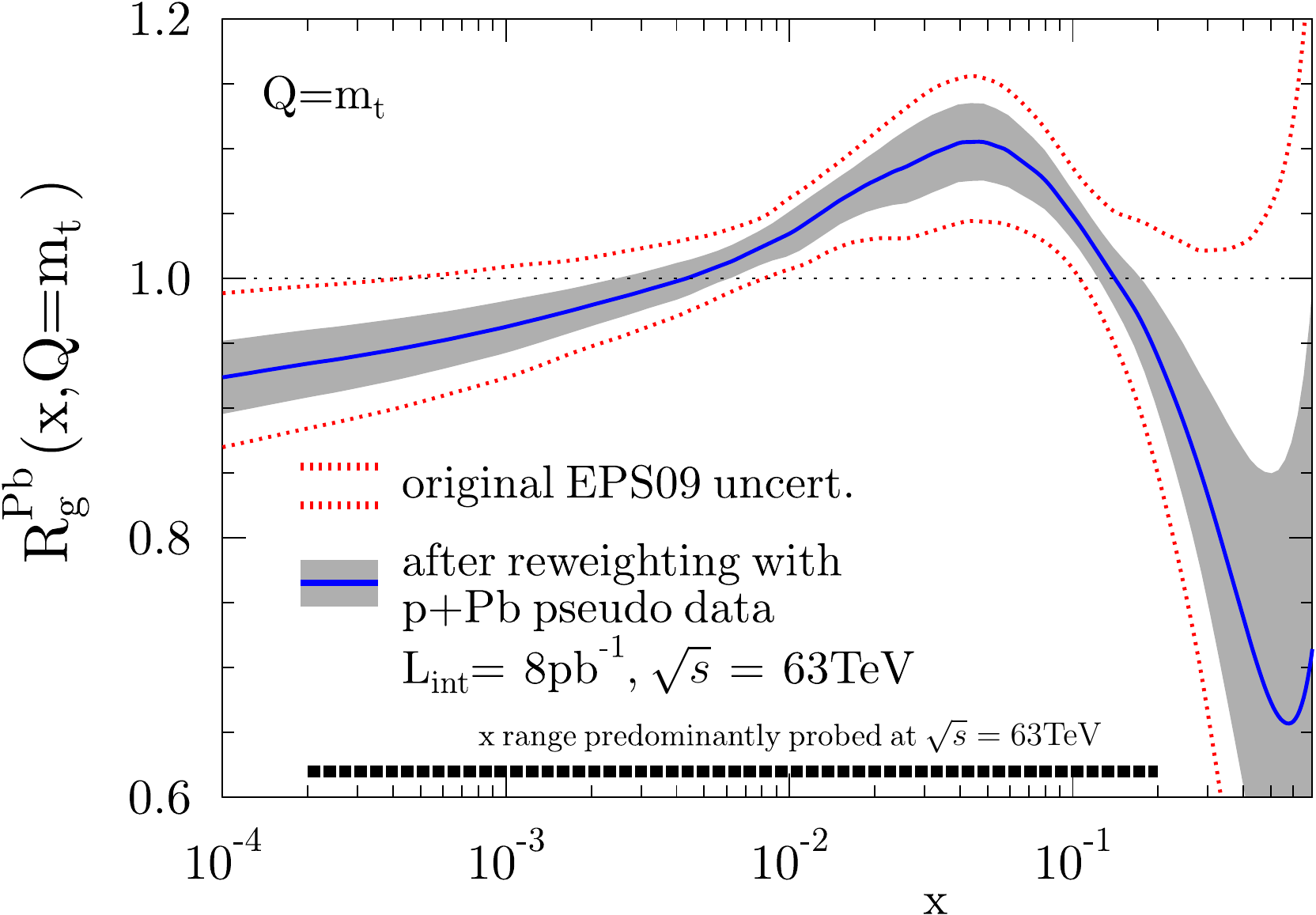}\hspace{0.1cm}
\includegraphics[width=0.48\textwidth]{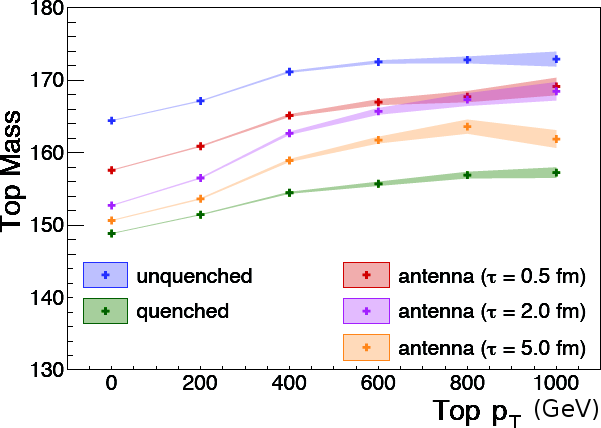}
\caption{Left: Ratio of Pb over p nuclear EPS09 gluon at $Q=m_{\rm top}$ (red dashes) and estimated improvements 
obtained using FCC pPb pseudodata (grey band)~\cite{d'Enterria:2015mgr,dEnterria:2017jyt}.
Right: Reconstructed top mass as a function of top-quark $\pT$ for various energy loss scenarios~\cite{top_Eloss}.}
\label{fig:2}
\end{figure*}

The top quark production cross sections at FCC energies are very large: $\sigma(\ttbar)\approx$~3.2,~300~$\mu$b 
in pPb and PbPb respectively~\cite{d'Enterria:2015mgr,dEnterria:2017jyt}, 
\ie\ $\times$55--85 larger than at the LHC. The corresponding yields per month, after typical analysis 
cuts, amount to few $10^5$ $\ttbar$ pairs in the ``cleanest'' fully leptonic decay channel, 
$\ttbar\to \bbbar\,2\ell\,2\nu$~\cite{dEnterria:2017jyt} (or about $\times$4 more in the lepton+jets modes). 
Top quarks provide valuable information of both 
the initial and final states in nuclear collisions. Since top pairs are dominantly produced in $gg$ fusion
processes, the rapidity distribution of their decay leptons in pPb compared to pp collisions traces accurately 
the nuclear modification of the gluon nPDF at high scales $Q = m_{\rm top}$ over a wide $x$ range~\cite{d'Enterria:2015mgr}.
The impact that the FCC $\ttbar$ data would have in a nPDF EPS09 global fit~\cite{Eskola:2009uj} 
is shown via the $R_{\rm g}^{\rm Pb}(x,Q^2) = g_{\rm Pb}(x,Q^2)/g_{\rm p}(x,Q^2)$ ratio
in Fig.~\ref{fig:2} (left). The uncertainties on the nuclear glue are reduced by about a factor
of two (grey band compared to red dashes), mostly in the so-far unexplored antishadowing and EMC $x$ 
regions~\cite{d'Enterria:2015mgr,Dainese:2016gch}.

The top quark can also be used as sensitive probe of in-medium parton energy loss by studying boosted 
topologies with increasingly longer (Lorentz-dilated) decays into final-state quarks $t\to W(\qqbar')b$.
Due to the Lorentz boost, the system travels through the medium as a colour-singlet antenna 
for some fraction of the time.
By comparing the reconstructed top mass as a function of its $\pT$ (Fig.~\ref{fig:2}, right), it should be thus possible 
to get a unique insight into the space-time evolution of jet-quenching in the hot QCD medium~\cite{top_Eloss}.

\section{Higgs boson suppression, and photon-photon collisions}

\begin{figure*}[hbpt!]
\centering
\includegraphics[width=0.45\textwidth,height=5.6cm]{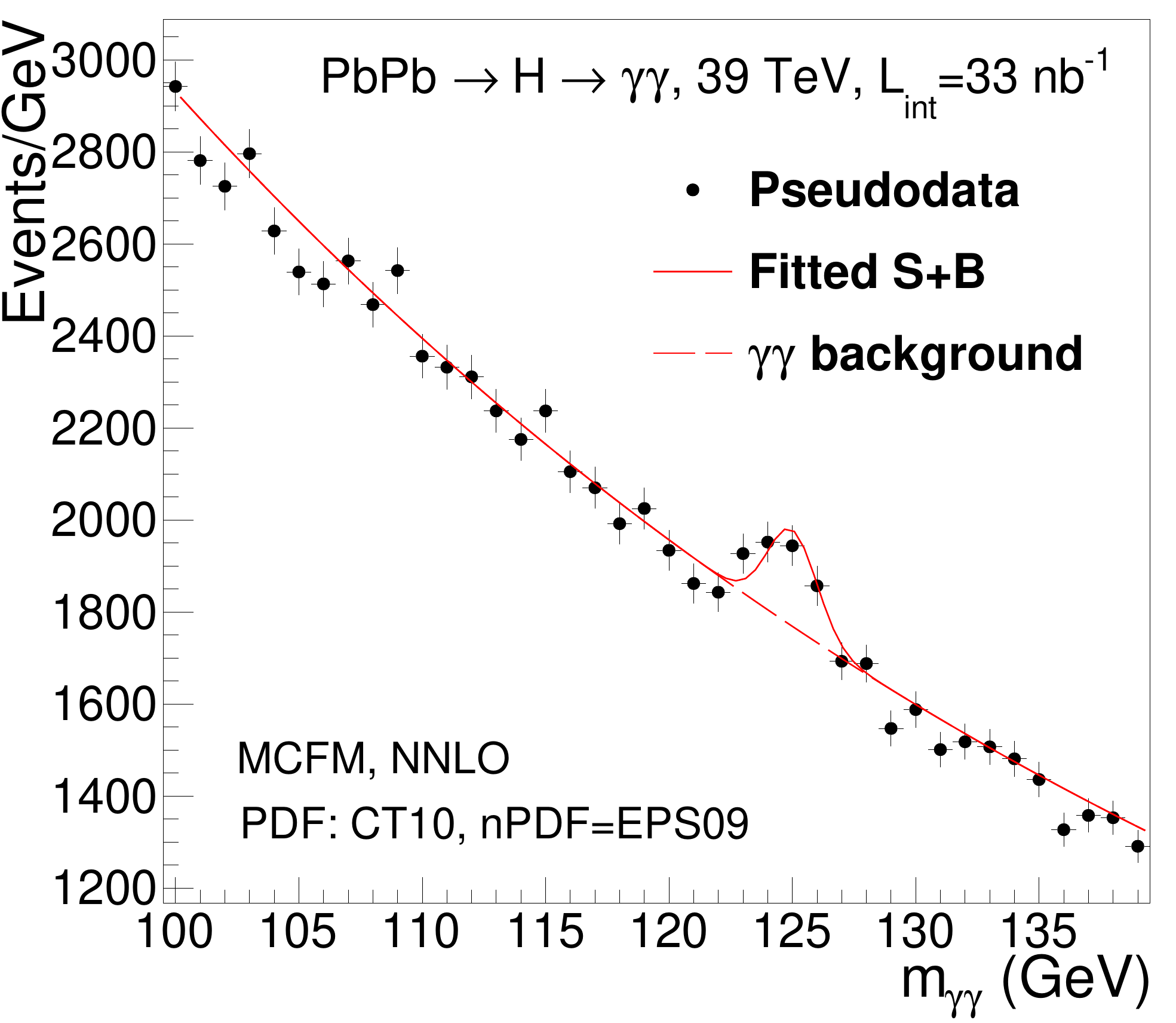}\hspace{0.0cm}
\includegraphics[width=0.54\textwidth,height=5.6cm]{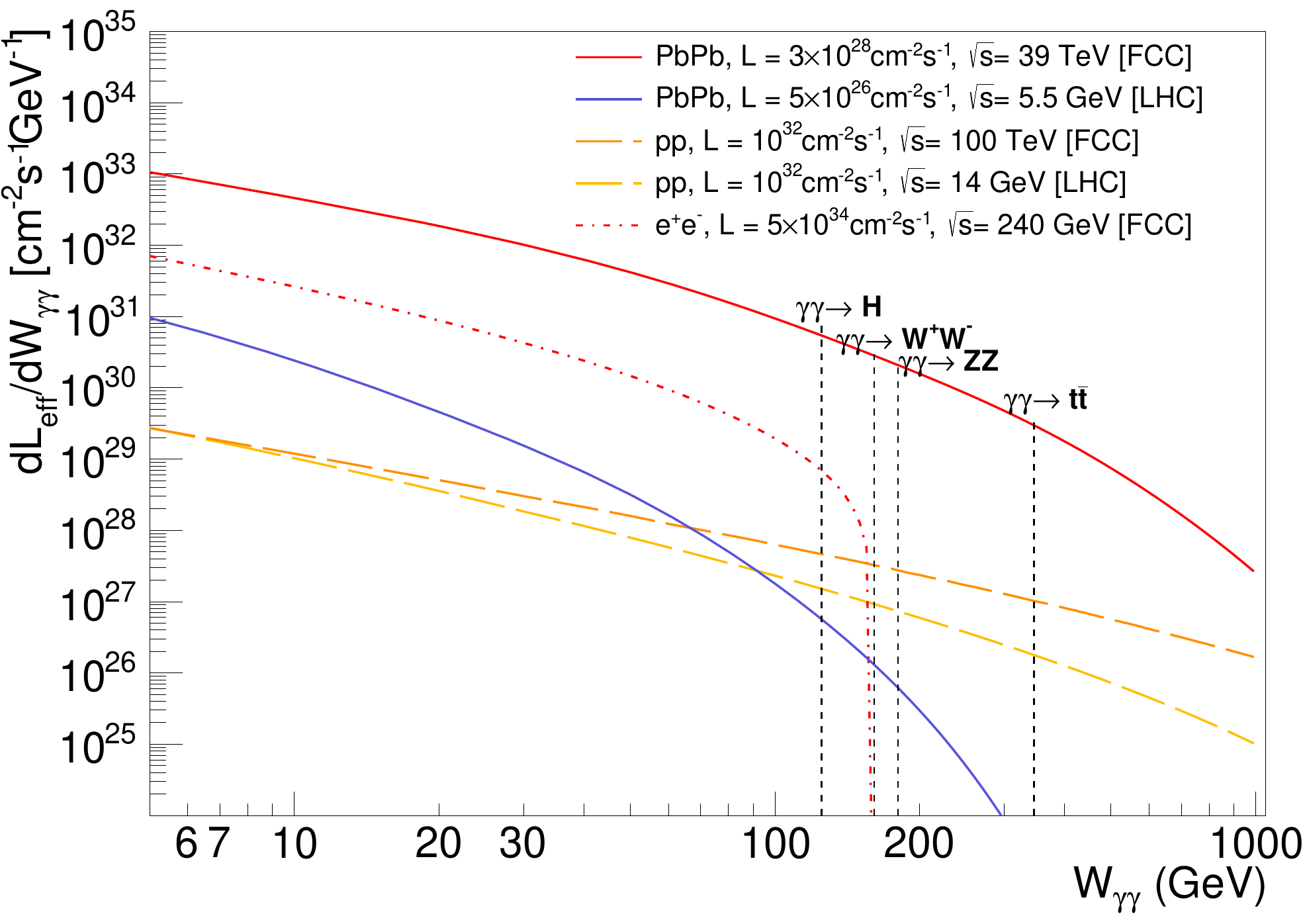}
\caption{Left: Expected diphoton invariant mass in PbPb at $\sqrtsnn$~=~39~TeV 
for a Higgs boson signal injected on top of the $\gaga$ backgrounds~\cite{dEnterria:2017jyt}.
Right: Effective photon-photon luminosities as a function of $\gaga$ \cm\ energy ($W_{\gaga}$)
for 5 different systems at FCC and LHC energies (the vertical dashed lines indicate the
thresholds for Higgs, $W^+W^-$, Z\,Z, and $\ttbar$ photon-fusion production.}
\label{fig:3}
\end{figure*}

The FCC will also allow for the observation of the Higgs boson in nuclear collisions since its production 
cross sections, $\sigma(gg\to \rm H)\approx$~0.12,~12~$\mu$b in pPb and PbPb respectively, are $\times$20 
larger than at the LHC~\cite{dEnterria:2017jyt}. In the clean diphoton ``discovery'' channel,  after branching 
ratio (BR~=~0.23\%), acceptance and efficiency losses, one expects about 500 and 1\,000 Higgs events per month 
in PbPb and pPb, on top of the corresponding $\gaga$ non-resonant background. Figure~\ref{fig:3} (left) shows the
expected invariant mass distribution in PbPb at 39~TeV. The expected significance of the diphoton signal 
(S) over the background (B), computed via S/$\sqrt{\rm B}$ at the Higgs peak, is 5.5$\sigma$.
Interestingly, once produced, the Higgs boson, with a lifetime of $\rm \tau_0 \approx$~50~fm/c, 
traverses the final-state medium and can scatter with the surrounding partons, resulting in a potential 
depletion of its yields compared to the pp case~\cite{DdE_CL}. Since the Higgs boson production cross
section in absence of final-state effects is theoretically very well known~\cite{Anastasiou:2015ema}, 
the amount of potential Higgs boson suppression can thus be used to accurately determine the density of the produced QGP.\\

Ultraperipheral interactions of ions, without hadronic overlap, provide very clean means to study photon-photon collisions 
at high energies, as their luminosities are extremely enhanced in nuclear compared to proton or electron beams 
(by a factor $Z^4 = 5\cdot10^7$ for PbPb)~\cite{Baltz:2007kq}. Figure~\ref{fig:3} (right) shows the effective 
photon-photon luminosity as a function of the $\gaga$ \cm\ energy reachable in various colliding systems 
($\epem$, pp, and PbPb) at the LHC and FCC energies. Clearly, PbPb at 39 TeV provides the largest two-photon 
luminosities, reaching for the first time the energy range beyond 1~TeV. The vertical lines show the thresholds for 
photon-fusion production of Higgs, $W^+W^-$, $ZZ$, and $\ttbar$. All such processes, sensitive to different tests 
of the electroweak sector of the Standard Model, will have visible counts at the FCC in its PbPb running mode.


\section{Summary}

The FCC provides unique physics capabilities for heavy-ion physics in pPb and PbPb collisions 
with $\times$7 and $\times$10 larger \cm\ energies and integrated luminosities than at the LHC. 
The expected particle and energy densities will be twice above those reached at the LHC, 
leading to medium temperatures in the 1~GeV range; charm densities will be $\times$3 those at the LHC, 
leading to a possibly enhanced $\jpsi$ production by recombination in the plasma; 
and the tightly-bounded $\Upsilon$(1S) resonance can be melted in the hot QGP. The large
\cm\ energies will open up studies of nuclear parton distributions in the unexplored
saturation (down to $x\approx$~10$^{-7}$) and large-$x$ (antishadowing
and EMC domains, with top-quarks) regimes. The heaviest SM elementary particles (top quark and Higgs boson) 
will be copiously produced, offering novel high-precision handles to study the properties of hot
and dense QCD matter. Last but not least, ultraperipheral nuclear collisions will provide high-luminosity
two-photon collisions at \cm\ energies reaching the TeV range. 


\end{document}